\documentclass[12pt]{article}
\usepackage{latexsym}
\usepackage{epsf}
\def\be{\begin{equation}}
\def\ee{\end{equation}}
\def\bea{\begin{eqnarray}}
\def\eea{\end{eqnarray}}

\def \alfa {2 \pi \alpha'}
\newcommand{\w}[1]{\\[0.#1cm]}

\def\be{\begin{equation}}
\def\ee{\end{equation}}
\def\ba{\begin{array}}
\def\ea{\end{array}}
\def\bea{\begin{eqnarray}}
\def\eea{\end{eqnarray}}

\def \unit{\hbox to 3.3pt{\hskip1.3pt \vrule height 7pt width .4pt \hskip.7pt
\vrule height 7.85pt width .4pt \kern-2.4pt
\hrulefill \kern-3pt
\raise 4pt\hbox{\char'40}}}

\def \x{\times}

\def \alfa {2 \pi \alpha'}
\def \j {{\sf g}} 

\begin{document}

\begin{flushright}
\footnotesize
UG-24/98\\
SPIN-98/16 \\
{\bf hep-th/9812225}\\
December, $1998$
\normalsize
\end{flushright}

\begin{center}


\vspace{.6cm}
{\Large {\bf Brane Actions and String Dualities}}\footnote{To be published
in the proceedings of the TMR-meeting
{\it Quantum Aspects of Gauge Theories, Supersymmetry and
Unification}, Corfu, Greece, September 1998.}

\vspace{.9cm}


{\bf Eduardo Eyras}

\vspace{.1cm}

{
{\it Institute for Theoretical Physics\\
University of Groningen \\
Nijenborgh 4, 9747 AG Groningen, The Netherlands}\\
{\tt E.A.Eyras@phys.rug.nl}
}

\vspace{.3cm}

{and}

\vspace{.3cm}

{\bf Yolanda Lozano}

\vspace{.1cm}

{
{\it Spinoza Institute\\
University of Utrecht\\
Leuvenlaan 4, 3508 TD Utrecht, The Netherlands}\\
{\tt Y.Lozano@phys.uu.nl}
}

\vspace{.2cm}


\vspace{.2cm}

\vspace{.8cm}


{\bf Abstract}

\end{center}

\begin{quotation}

\small

An effective action for the M9-brane is proposed.
We study its relation with other branes via dualities. 
Among these, we find actions for branes which are 
not suggested by the central charges of the Type II superalgebras.

\end{quotation}

\vspace{1cm}

\newpage

\pagestyle{plain}


\newpage
\section{Introduction}

The BPS states give information about the non-perturbative structure
of M- and string theory. There is a known relation
between the central charge structure of the spacetime
supersymmetry algebra and the occurrence of BPS states in the theory.
The generic rule is that the presence of a $p$-form central charge in 
$D$ dimensions suggests the existence of a $p$-brane and a ($D-p$)-brane 
\cite{to1,Hull}. The eleven-dimensional supersymmetry
algebra with the maximum number of central charges
is given by ($\alpha = 1,\dots,32; M=0,\dots,10$):
\begin{eqnarray}
\{Q_\alpha, Q_\beta\} &=& \left (\Gamma^MC\right)_{\alpha\beta}P_M
+ {\textstyle {1\over 2!}}\left (\Gamma^{MN}C\right)_{\alpha\beta}
Z_{MN}  \nonumber
\w2
&&+{\textstyle{1\over 5!}}\left (\Gamma^{MNPQR}C\right )_{\alpha\beta}
Z_{MNPQR}\, . 
\end{eqnarray}
The spatial component of $P_M$ is related to an M-Wave, moreover we find
an M2-brane, an M5-brane, an MKK-monopole and an M9-brane:
\be
Z_{MN} \rightarrow  {\rm M2 \,\, and \,\, M9} \, ,\qquad
Z_{MNPQR} \rightarrow  {\rm M5 \,\, and \,\, MKK} \, .
\ee
The dynamics of the M2- and M5- branes have been extensively studied.
However, it is only recently that the dynamics of the Kaluza-Klein
monopole has been understood \cite{Hull,BJO}. The nature of the M9-brane
is not yet well understood. 

One can perform a similar analysis of 
the Type IIA supersymmetry algebra with
the maximal number of  central charges. 
One then finds a gravitational wave (W-A), a
fundamental string (NS-1A), D$p$-branes ($p = 0,2,4,6,8$), a solitonic
five-brane(NS-5A), a Kaluza-Klein monopole (KK-5A) and 
a nine-brane (NS--9A). 
Similarly, from the Type IIB supersymmetry algebra
one finds a gravitational wave (W-B), a fundamental string
(NS-1B), D$p$-branes ($p = 1,3,5,7,9$), a solitonic five-brane (NS-5B), a
Kaluza-Klein monopole (KK-5B) and a further nine-brane (NS-9B).
We see that in particular
an S-dual partner for the D9-brane is predicted \cite{Hull}.
The role of these new 9-branes in string theory is not yet well
understood (see however \cite{BEHHLvdS}).

All these branes are related to each other by T- and S- dualities,
and the IIA branes can be obtained from the M-branes by dimensional
reduction.
In this contribution we review the dynamics of the MKK-monopole
and propose an effective action for an M9-brane in similar terms.
Studying their relation with other branes via dualities, we find 
branes which are not suggested by the supersymmetry algebras.


\subsection{The M-theory KK-monopole}


The Kaluza-Klein monopole is a purely gravitational solution
that appears in Kaluza-Klein compactification \cite{Gross-Perry}.
In 11 dimensions, it can be schematically represented by\footnote{We use
a notation where $\times$ ($-$) denotes a worldvolume
(transversal) direction and $z$ corresponds to the isometry direction.}:
\be
 {\rm MKK-monopole} :
\mbox{ 
$\left\{ \begin{array}{c|cccccccccc}
           \x &\  \x  & \x  & \x & \x & \x & \x & z & - & - & -            
                                    \end{array} \right.$}    \, .
\ee
It has 8 isometry directions, one of which is
generated by a Killing vector\footnote{We use 
hats for 11-dimensional fields and no hats for 10-dimensional
ones.} ${\hat k}=\partial_z$. Thus 
it can be considered as an extended object. 
The worldvolume field content must correspond, after gauge fixing, 
to a 7-dimensional vector multiplet \cite{Hull} with 3 scalars and 1 vector.
However, if we consider the isometry direction $z$ either as worldvolume
or transversal, we do not obtain the right number of degrees of freedom.
In order to solve this, we use a characteristic feature of the monopole. 
Since the monopole is localized in the transversal Taub-NUT 
space, the $z$ isometry must be perpendicular 
to the worldvolume of the monopole action.
However, we have in general ${\hat k}_M \partial_i {\hat X}^M \neq 0$,
for embedding coordinates ${\hat X}^M$ and worldvolume coordinates $\xi^i$.
We define then a projector:
\be
{\cal P}_M{}^N = \delta_M{}^N + |{\hat k}|^{-2} {\hat k}_M {\hat k}^N \, ,
\ee
which projects any vector to the component perpendicular to ${\hat k}$.
The monopole is then described by the embeddings
$\partial_i {\hat X}^M {\cal P}_M{}^N$, which satisfy
${\hat k}_N \partial_i X^M {\cal P}_M{}^N = 0$.

With this construction we have gauged the
translations along ${\hat k}$,
since the 
projected pullback is in fact equivalent to a covariant derivative:
\be
D_i{\hat X}^M = \partial_i {\hat X}^M + {\hat A}_i {\hat k}^M \, ,
\ee
with gauge field $
{\hat A}_{i}= |{\hat k}|^{-2} \partial_{i} {\hat X}^{M} {\hat k}_{M}$.
Therefore, after gauge fixing, 
the embedding scalar ${\hat X}^z$ disappears. Consequently we achieve the 
right number of degrees of freedom.
The kinetic term of the action has the following form \cite{BJO}:
\be 
{\hat S} = - T_{\rm MKK} \int d^7 \xi \,\, |{\hat k}|^2 
\sqrt{|{\rm det}
\left({\hat \Pi} + l_p^2 |{\hat k}|^{-1}{\hat {\cal K}}^{(2)}\right)|} \, ,
\label{MKK-action}
\ee
where the field-strength ${\hat {\cal K}}^{(2)}$
of the Born-Infeld-like 1-form ${\hat \omega}^{(1)}$ is defined by
\begin{equation}
\label{efe}
\begin{array}{rcl}
{\hat {\cal K}}^{(2)} &=& d {\hat \omega}^{(1)} + l_p^{-2}
\, \partial{\hat X}^{M} \partial{\hat X}^{N} 
(i_{\hat k} {\hat C}^{(3)})_{MN} \, , \\
\end{array}
\end{equation}
and ${\hat \Pi}$ is the pullback of the spacetime metric:
\be
{\hat \Pi} = D {\hat X}^{M} D {\hat X}^{N} {\hat g}_{{M}{N}} = 
\partial {\hat X}^M \partial {\hat X}^N \left( {\hat g}_{MN}
+ |{\hat k}|^{-2}{\hat k}_M{\hat k}_N \right) \, .
\ee
In general, for all KK-monopoles, there is a coupling to the
Killing vector through the scalar
\be
|{\hat k}|^2 = \left( {R_k \over l_p} \right)^2 \, ,
\ee
where $R_k$ is the radius of the compact isometry and
$|{\hat k}|^2 = - {\hat k}^{M}{\hat k}^M {\hat g}_{MN}$.
Accordingly, the tension of the 
KK-monopole has the following form\footnote{We use
conventions for which $G_{11} = 16\pi^7 l_p^9$.}:
\be
{\cal T}_{\rm MKK} = {R_k^2 \over (2\pi)^6l_p^9} \, .
\ee
The complete action for the KK-monopole in different
backgrounds has been given in \cite{Bert,BEL,EJL}.
The 1-form field ${\hat \omega}^{(1)}$ 
describes the flux of an M2-brane wrapped around the isometry
direction. This corresponds to the  intersection 
of an M2-brane with the MKK-monopole over a 0-brane, such that one of the
worldvolume directions of the M2-brane coincides with the
isometry direction $z$ of the Taub-NUT space \cite{Groningen-Boys-2}.
A reduction of this intersection 
along the $z$--direction gives the configuration $(0|{\rm F}1,{\rm D}6)$. 
The reduction of ${\hat \omega}^{(1)}$ leads in this case
to the BI field of the  D6-brane, describing the tension of 
the fundamental string. 


\subsection{The M9-brane}


The M9-brane is believed to be the strong coupling 
limit of the D8-brane \cite{BdRGPT}.
However, the D8-brane is only consistent in the context of
a Type ${\rm I}^\prime$ construction \cite{Pol-2}, where 
there are 2 orientifold 8-planes at the ends of one
interval and 32 D8-branes located along the interval. 
In general, the background between two D8-branes
is given by massive IIA supergravity. 
On the other hand, it is known \cite{PW}
that only in the case where 16 D8-branes are sitting
on top of each orientifold, there exists a strong coupling limit
where the theory becomes eleven-dimensional. In this situation,
the background in the bulk is standard IIA supergravity.
This is related to the fact that massive IIA supergravity can only
be given an eleven dimensional interpretation in terms of
compactified M-theory \cite{BLO}.


It is then natural to consider an M9-brane as part of a composite
of sixteen M9-branes and one M-theory orientifold plane,
given that an M9-brane cannot be singled out in 
eleven uncompactified dimensions.
As discussed in \cite{BEHHLvdS} this system would describe the
{\it end of the world} branes of Ho{\v r}ava and Witten 
\cite{Horava-Witten-2}.

If we consider one compact direction, we can
single out one of the M9-branes from the composite,
trading the Wilson lines of the heterotic theory for positions
of the M9-branes \cite{Joe's-book,BEHHLvdS}. The background between 
the M9-branes is the {\it massive} 11-dimensional supergravity of \cite{BLO}.
Accordingly, the dynamics of a single M9-brane will have associated
a Killing isometry describing this compact direction,
which means that it will be  
naturally described by a gauged 
sigma-model\footnote{This was also suggested in \cite{BJO,BvdS}, 
based on other arguments.}.
According to this  discussion, an M9-brane
can be represented schematically by:
\begin{displaymath}
 {\rm M}9-{\rm brane} :
\mbox{ 
$\left\{ \begin{array}{c|cccccccccc}
           \x & \  z & \x  & \x & \x & \x & \x & \x & \x & \x & -            
                                    \end{array} \right.$}    \, ,
\label{M9-brane-schematically}
\end{displaymath}
where $z$ indicates the direction of the Killing isometry.
We propose the following gauged sigma-model to describe 
the dynamics of a single M9-brane:
\begin{equation}
{\hat S} = - T_{\rm M9} \int d^9 \xi \,\, |{\hat k}|^3 
\sqrt{|{\rm det} \left( {\hat \Pi}+ |{\hat k}|^{-1} l_p^2 
{\hat {\cal K}}^{(2)} \right)|} \, ,\\
\label{M9-action}
\end{equation}
\noindent with ${\hat \Pi}$ and ${\hat {\cal K}}^{(2)}$
defined as for the MKK-monopole.
Notice that, strictly speaking, this M9-brane is actually 
an 8-brane with one extra isometry. The M9-brane worldvolume corresponds
to a 9-dimensional vector multiplet with one vector and one scalar,
where one transversal scalar has been eliminated by the gauging.
The action is such that dimensionally reducing along the
Killing vector gives the D8-brane action. 
The effective tension of the M9-brane can be written as
\cite{Hull-M-theory}:
\be
{\cal T}_{\rm M9}= {R_k^3 \over (2\pi)^8 l_p^{12}} \, .
\ee
Similarly to the MKK-monopole case, 
the 1-form field ${\hat \omega}^{(1)}$ 
appearing in the M9-brane action describes the
flux of an M2-brane which has one direction wrapped around the
isometry direction $z$. This corresponds to the intersection of an
M2-brane with an M9-brane, which gives upon reduction
along the isometry direction 
the intersection of a fundamental string with a D8-brane.


\section{Gauged sigma-models and Type II dualities}


Since the MKK-monopole and the M9-brane have similar actions, we can 
describe the reduction to Type IIA branes 
in a uniform fashion for both branes.
The reduction of the gauged sigma-models
(\ref{MKK-action}) and (\ref{M9-action}) along the Killing isometry gives
a D$p$-brane. A D6-brane from the MKK and a D8-brane from the M9.
Any other type of reduction gives again a gauged sigma-model
in Type IIA. 
In general, the Killing vector ${\hat k}$ becomes, after reduction along 
an eleventh-coordinate $Y$,
the Killing vector associated to the isometry of the reduced brane:
${\hat k}^y = 0$, ${\hat k}^\mu = k^\mu$.
We consider first the double dimensional reduction of the
MKK/M9 action. We find\footnote{We use the notation of \cite{EJL}.}:
\begin{eqnarray}
\lefteqn{ S =
- \int d^{p+1} \xi \,\,
e^{-{p-1 \over 2}\phi} |k|^{p-1 \over 2} 
\left( 1 + e^{2 \phi}|k|^{-2} (i_k C^{(1)})^2 \right)^{p-3 \over 4} \times}
\\
&&\times\sqrt{|{\rm det}(\Pi_{ij} 
- (\alfa)^2 |k|^{-2} {\cal K}^{(1)}_i {\cal K}^{(1)}_j +
 { \alfa  |k|^{-1} e^\phi \over \sqrt{ 1 + e^{2 \phi}|k|^{-2} 
(i_k C^{(1)})^2}} {\cal K}^{(2)}_{ij})|} \, .\nonumber
\label{KKA-7A-action}
\end{eqnarray}
This action corresponds to the KK-5A monopole for $p=5$.
For $p=7$ we call it a KK-7A brane.
The field-strengths are defined as follows:
\begin{equation}
\ba{rcl}
{\cal K}^{(1)} &=& d \omega^{(0)} - {1 \over 2\pi\alpha^{\prime}}
(i_k B) \, ,\\& &\\
{\cal K}^{(2)} &=& d \omega^{(1)} + {1 \over 2\pi\alpha^{\prime}}
(i_k C^{(3)}) - {\cal K}^{(1)} \wedge DX^\mu C_{\mu}^{(1)} \, ,
\ea
\end{equation}
where $B$ is the NS-NS 2-form and $C^{(p)}$ denotes a R-R $p$-form.
The covariant derivatives are defined as usual.
The effective tension for these branes is given by:
\be
{\cal T}_p = {R_k^{{p-1\over 2}} \over (2\pi)^p g_s^{p-1 \over 2}
l_s^{3p+1 \over 2}} \, ,\qquad p=5,7 \, .
\ee
On the other hand, the reduction 
of the MKK/M9 along one transversal direction
gives rise to a worldvolume scalar $c^{(0)} = (\alfa)^{-1} Y$. 
The action for this brane is given by
\begin{eqnarray}
\lefteqn{ S = - \int d^{p+1} \xi \,\, e^{-{p \over 2} \phi} 
|k|^{{p\over 2} -1}
\left( 1 + |k|^{-2} e^{2 \phi} (i_k C^{(1)}) \right)^{p-2 \over 4} \times}
\\
&& \hspace{-.5cm}\times \sqrt{|{\rm det}
(\Pi_{ij} 
- {(\alfa)^2 e^{2\phi} \over 1+ |k|^{-2} e^{2\phi} (i_k C^{(1)})^2}
{\cal G}_i^{(1)}{\cal G}_j^{(1)} + {\alfa |k|^{-1} e^\phi \over 
\sqrt{1+ |k|^{-2} e^{2\phi} (i_k C^{(1)})^2}}
{\cal H}_{ij}^{(2)} )|} \, .\nonumber
\label{6A-9A-action}
\end{eqnarray}
For $p=8$ it corresponds to the NS-9A brane (as shown explicitly
in \cite{BEHHLvdS}).
We call it here a KK-8A brane to make clear that it contains
a gauged isometry. For $p=6$, it gives a new
brane which is not predicted by the Type IIA supersymmetry algebra.
We call it a KK-6A brane.
The field-strengths are now defined as follows:
\begin{equation}
\begin{array}{rcl}
{\cal G}^{(1)} &=& d c^{(0)} + {1 \over \alfa} DX^\mu C^{(1)}_\mu \, ,
\\& &\\
{\cal H}^{(2)} &=& d v^{(1)} + {1 \over \alfa}(i_k C^{(3)})
- (i_k B) \wedge d c^{(0)} \, ,\\
\end{array}
\end{equation}
The fields $\omega^{(1)}$ and $v^{(1)}$ have the interpretation of the
flux of a D2-brane ending on the corresponding gauged sigma-model.
The effective tension for the KK-6A/KK-8A branes is given by:
\be
{\cal T}_p = {R_k^{{p\over 2}-1} \over (2\pi)^p g_s^{p \over 2}
l_s^{{3\over 2}p}} \, ,\qquad p=6,8 \, .
\ee
At the level of the solutions of the Type II supergravities
it is known that
T-duality relates the KK-monopole with the NS-5 brane, where the duality
is performed in the isometry direction of the monopole and
a transversal direction of the 5-brane.
This also works at the level of the worldvolume effective actions.
Then one can construct the effective action of the 
NS-5B brane by applying T-duality to the KK-A monopole action.
The result is exactly the S-dual of the D5-brane action \cite{EJL}.
Similarly, applying T-duality to the action of the NS-5A brane, one
finds the action for the Type IIB KK-monopole.
The solutions corresponding to the KK-6A and KK-7A branes
and their relations with other branes via dualities 
have been considered in \cite{PaTo}.

\begin{figure}[!ht]
\begin{center}
\leavevmode
\epsfxsize= 12.5cm
\epsfysize= 4cm
\epsffile{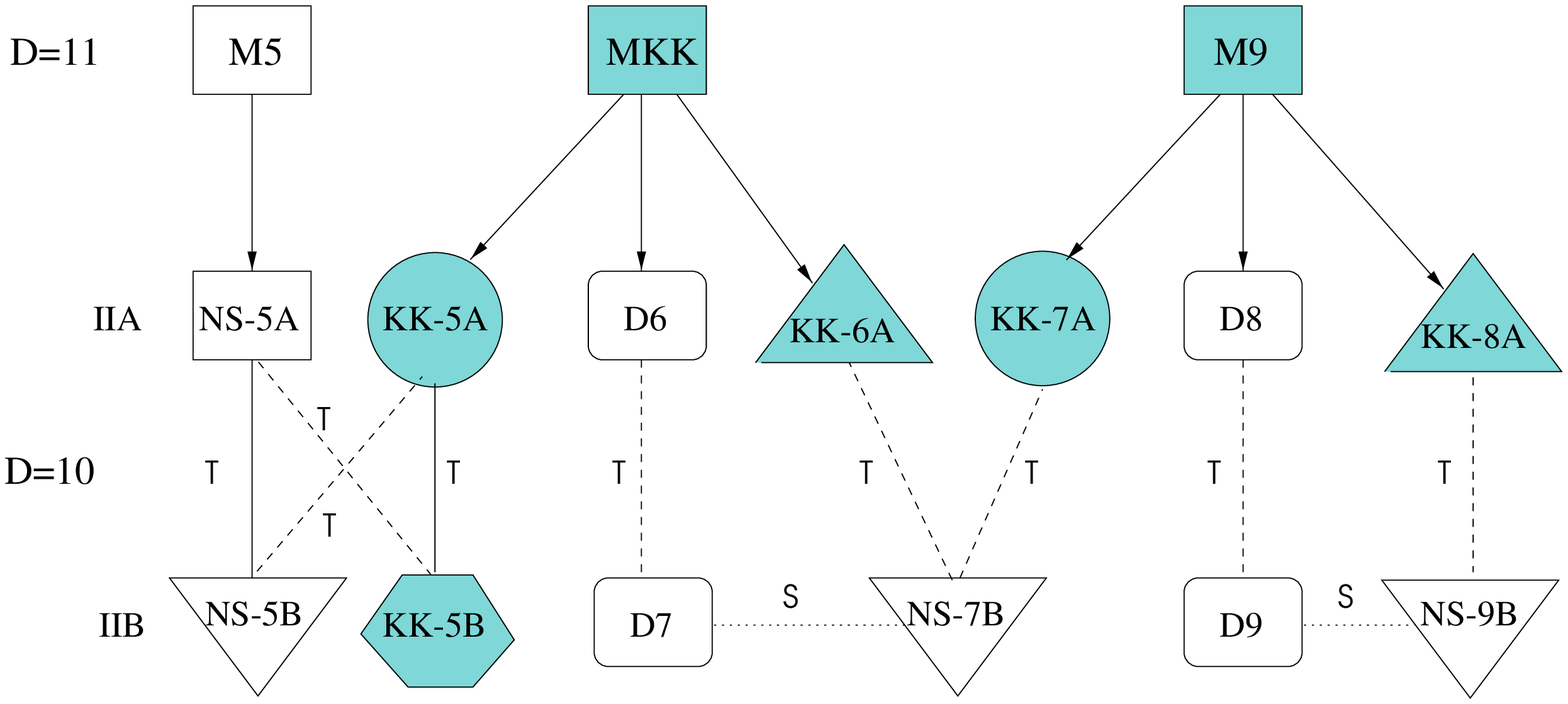}
\caption{\footnotesize {\bf Gauged sigma-models and dualities.}
This figure depicts the duality relations
between the Type IIA branes obtained by reduction of the 
MKK-monopole and M9-brane and other branes in Type IIB.
The branes described by a gauged sigma-model are represented in grey.
Branes with similar worldvolume theories are represented by the same
geometrical figure.}
\label{fig:KKA-6A-7A-9A}
\end{center}
\end{figure}

Here we are going to show that the  
KK-$p$A branes described by
(\ref{KKA-7A-action}) and (\ref{6A-9A-action}) are
in general T-dual to the solitonic $p$-branes which are
S-dual partners of the Type IIB D$p$-branes ($p$ odd and $p>3$)
(see Figure \ref{fig:KKA-6A-7A-9A}), whose actions are  given by \cite{EJL}:
\begin{equation}
\int d^{p+1}  \xi \,\, e^{- ( {p-1 \over 2})\varphi }
( 1 + e^{2 \varphi} C^{(0) \, 2})^{ p-3 \over 4}
\sqrt{| {\rm det} \left( \j + { (\alfa) e^\varphi \over
\sqrt{1+e^{2\varphi}C^{(0) \, 2}}} {\tilde {\cal F}} \right)|} \, .
\label{hartura}
\end{equation}
This action 
is obtained by directly applying S-duality on the D$p$-brane action.
The curvature ${\tilde {\cal F}}$ is given by:
\be
{\tilde {\cal F}} = dc^{(1)} + {1 \over \alfa} C^{(2)} \, ,
\ee
where the 1-form $c^{(1)}$  is a Born-Infeld-like field
representing the flux of a D-string ending on the brane
\cite{Townsend}.
The effective tension for these branes is given by:
\be
{\cal T}_p = {1 \over (2\pi)^p g_s^{p-1 \over 2} l_s^{p+1}} \, ,\qquad
p=5,7,9 \, .
\ee
The action (\ref{hartura}) represents a
NS-5B brane for $p=5$, as derived in \cite{EJL}.
For $p=7$, it corresponds to a NS-7B. However, this is not
predicted by the Type IIB supersymmetry algebra and 
indeed there seems
to be only one seven-brane in Type IIB
\cite{Gibbons-Green-Perry,Hull}.
It is not yet clear what the role of this NS-7B brane is.
Finally, for $p=9$ it gives the action of the NS-9B brane 
\cite{BEHHLvdS}. 

T-duality on the gauged sigma-models
(\ref{KKA-7A-action}) and 
(\ref{6A-9A-action}) is performed along their Killing direction. 
For the Type IIB branes (\ref{hartura}), the T-duality
is performed along a transversal direction when it connects them to
the KK-5A and KK-7A branes. However,
for the case of the KK-6A and KK-8A branes,
the T-duality is performed along a worldvolume direction.
For instance, the KK-5A/KK-7A are T-dual to the NS-5B/NS-7B branes,
respectively.
The relation between the worldvolume fields is given by:
\be
\omega^{(0) \prime} = {1 \over \alfa} Z \, ,\qquad
\omega^{(1) \prime} = - c^{(1)} \, ,
\ee
where $Z$ represents a transversal embedding coordinate of the
NS-5B/NS-7B brane.
On the other hand, the KK-6A/KK-8A are T-dual to the 
NS-7B/NS-9B branes, respectively. 
In this case the T-duality rules are given by
\be
v^{(1) \prime}_i = - c^{(1)}_i \, ,
\qquad c^{(0) \prime} = - c^{(1)}_\sigma \, ,
\ee
where $\sigma$ is the wrapped worldvolume direction of the NS-7B/NS-9B brane.

The Type II supersymmetry algebras also predict the presence of two 
9-branes in Type IIB. These  9-branes 
and the D8- and KK-8A 
branes are related by a chain of S- and T- dualities
(See Figure \ref{fig:KKA-6A-7A-9A}).
The D8-brane is known to be the T-dual of the D9-brane.
On the other hand, by the IIB supersymmetry algebra
there is an S-dual partner of the D9-brane, the NS-9B brane \cite{Hull},
which is  given by (\ref{hartura}) for $p=9$.
This 9-brane is T-dual to the KK-8A brane.
Accordingly, the Killing isometry has the interpretation of an
eleventh circular coordinate, from the point of view of the D8-branes;
or a tenth circular coordinate, from the point of view of the 
KK-8A brane \cite{BEHHLvdS}.

The D8- and D9- branes play a role in the orientifold constructions
of the Type IIA and IIB theories.
The duality relations above suggest that the NS-9B and KK-8A branes
should play a similar role in the duality related theories
(this is extensively studied in \cite{BEHHLvdS}).



\section{Conclusions}

We have presented a dynamical description for a single M9-brane
in terms of a gauged sigma-model, similar to 
that describing the MKK-monopole.
The descendants of these two branes have been considered in a unified fashion,
with the result that two of them are not predicted  by the 
Type IIA supersymmetry algebra: 
the KK-6A and KK-7A branes. 
The KK-$p$ branes for $p=6,7,8$, have an exotic 
dependence in the string coupling,
proportional to  $1/g_s^3$ and $1/g_s^4$. 
Thus they do not have an obvious interpretation 
in weakly coupled string theory, where the most singular behaviour
is expected to be $1/g_s^2$. 
Nevertheless, U-duality studied at the algebraic
level requires these extra states in order to fill up
multiplets of BPS states in representations of the U-duality symmetry group
of M-theory on a d-torus \cite{Hull-M-theory,Blau-O}.
These KK-$p$A branes are typically T-dual to 
the NS-$p$B branes defined by (\ref{hartura}).

The description of the M9-brane is not yet complete, since we are forced to
have a compactified 11-dimensional background.
However, one expects that there is a formulation for which,
putting 16 M9-branes together, one can get rid of 
the gauged isometry and go over to the {\it ends of the world}
description of \cite{Horava-Witten-2}. 


\section*{Acknowledgments}

We would like to thank A.~Ach\'ucarro, E.~Bergshoeff, R.~Halbersma, 
C.M.~Hull, N.~Obers, J.P.~van der Schaar, 
and especially D.~Matalliotakis for useful discussions.
E.E.~also thanks Nordita and Niels Bohr Institute 
for hospitality, and the organizers of the TMR-network Conference
{\it Quantum aspects of Gauge theories, Supersymmetry and Unification},
in Corfu, for giving him the opportunity to present this work.
The work of E.E. is part of the research program of the FOM.


\end{document}